\pdfoutput=1
\documentclass[
nofootinbib,longbibliography,
notitlepage,
aps]{revtex4-2}
\usepackage[margin=1in]{geometry}
\usepackage{amsmath,amsfonts,amssymb,amsthm,amsbsy,mathtools}
\usepackage{graphicx}
\usepackage{dcolumn}
\usepackage{bm}
\usepackage{physics} 
\usepackage{bbold}
\usepackage{mathtools}
\usepackage{xcolor}
\usepackage{tikz}
\usepackage{tikz-cd}
\usepackage{comment}
\usepackage{subfigure}
\usepackage{comment}
\usepackage{soul}
\usepackage{framed}
\usepackage{mdframed}
\usepackage{csquotes}
\usepackage{appendix}
\usepackage{hyperref}

\usepackage{natbib}
\usepackage[normalem]{ulem}
\usepackage{xcolor}
\usepackage{xspace}
\usepackage{ifthen}
\usepackage{wrapfig}

\newcommand{\showcomments}{true}

\newcommand{\acd}[1]%
{\ifthenelse{\equal{\showcomments}{true}}%
{{\color{magenta}{#1}}}{\xspace}}%

\newcommand{\vk}[1]%
{\ifthenelse{\equal{\showcomments}{true}}%
{{\color{orange}{#1}}}{\xspace}}%

\newcommand{\new}[1]%
{\ifthenelse{\equal{\showcomments}{true}}%
{{\color{black}{#1}}}{\xspace}}%

\newcommand{\neww}[1]%
{\ifthenelse{\equal{\showcomments}{true}}%
{{\color{red}{#1}}}{\xspace}}%

\newcommand{\strike}[1]%
{\ifthenelse{\equal{\showcomments}{true}}%
{{\color{gray}{\sout{#1}}}}{\xspace}}%

\begin{document}
\title{What an event is not: unravelling the identity of events in\\ quantum theory and gravity}

\author{Anne-Catherine de la Hamette$^{\ast,\dagger}$}
\affiliation{University of Vienna, Faculty of Physics, Vienna Doctoral School in Physics, and Vienna Center for Quantum Science and Technology (VCQ), Boltzmanngasse 5, A-1090 Vienna, Austria}
\affiliation{Institute for Quantum Optics and Quantum Information (IQOQI),
Austrian Academy of Sciences, Boltzmanngasse 3, A-1090 Vienna, Austria}

\author{Viktoria Kabel$^{\dagger}$}
\affiliation{University of Vienna, Faculty of Physics, Vienna Doctoral School in Physics, and Vienna Center for Quantum Science and Technology (VCQ), Boltzmanngasse 5, A-1090 Vienna, Austria}
\affiliation{Institute for Quantum Optics and Quantum Information (IQOQI),
Austrian Academy of Sciences, Boltzmanngasse 3, A-1090 Vienna, Austria}

\author{\v{C}aslav Brukner}
\affiliation{University of Vienna, Faculty of Physics, Vienna Doctoral School in Physics, and Vienna Center for Quantum Science and Technology (VCQ), Boltzmanngasse 5, A-1090 Vienna, Austria}
\affiliation{Institute for Quantum Optics and Quantum Information (IQOQI),
Austrian Academy of Sciences, Boltzmanngasse 3, A-1090 Vienna, Austria}

\thispagestyle{empty}
\begin{abstract}
\vspace*{1cm}
We explore the notion of events at the intersection between quantum physics and gravity, inspired by recent research on superpositions of semiclassical spacetimes. By going through various experiments and thought experiments -- from a decaying atom, to the double-slit experiment, to the quantum switch -- we analyse which properties can and cannot be used to define events in such non-classical contexts. Our findings suggest an operational, context-dependent definition of events which emphasises that their properties can be accessed without destroying or altering observed phenomena. We discuss the implications of this understanding of events for indefinite causal order as well as the non-absoluteness of events in the Wigner's friend thought experiment. These findings provide a first step for developing a notion of event in quantum spacetime.
\end{abstract}

\maketitle

\vspace{13mm}

\noindent \textit{Essay written for the Gravity Research Foundation 2024 Awards for Essays on Gravitation.}

\vspace{13mm}

\text{\hfill \hspace*{104mm} Submitted on March 29th, 2024}

\vfill

\noindent \rule{1cm}{0.2mm}

\noindent {\footnotesize $^{\ast}$Corresponding author: annecatherine.delahamette@univie.ac.at}

\noindent {\footnotesize $^{\dagger}$These authors contributed equally to this work.}
\newpage

\pagenumbering{arabic} 

Imagine yourself as an experimentalist capable of achieving the seemingly unachievable: placing planet Earth in a superposition of locations. Arguably, such a mass configuration would source a gravitational field in quantum superposition \cite{ChapelHill, delaHamette2021falling, aspelmeyer2021}. More concretely, we would expect to find the gravitational field in a quantum state peaked around two semiclassical configurations, one in which the source is on the left and one in which it is on the right \cite{Christodoulou_2019}. Now picture this: amidst this extraordinary experimental setup, two satellites orbit in opposite directions. Due to a slight oversight in orbital calibration, they collide, unleashing a catastrophic explosion. How would one describe such an event in a non-classical spacetime? Does it happen at the same spacetime point across the superposition, i.e.~is it localised in spacetime or not? This question is complicated by the fact that general relativity is a diffeomorphism-invariant theory. In this context, what we mean by an event happening at the same or different points across the superposition becomes ambiguous. We cannot simply \enquote{label} the spacetime points in each of the branches and identify those that carry the same label. The diffeomorphism invariance of general relativity provides us with the freedom to arbitrarily reshuffle the labels, that is, allows to freely change the coordinates. In particular, invoking the linearity of quantum theory, there is nothing that prevents us from changing the coordinates independently in each branch of the superposition. As a result, how we identify spacetime points across the branches of the superposition -- what constitutes the same or different spacetime points -- is not absolute \cite{kabel2024identification}. Thus, asking whether an event happens at the same or different spacetime points across the superposition does not have an absolute answer. However, if the localisation of the event does not provide an inherent description thereof, this prompts the question which properties do. Is it still the same event if the satellites collide at different velocities in the two branches? What if the collision occurs at a different angle or with a different force? What if the impact in one branch is not sufficient to cause an explosion in the first place? Is it, then, still the same event?

In this essay, we thus want to examine the question of which properties enable us to identify or distinguish events across different branches of a superposition. In fact, as we will see below, this question not only arises in the context of non-classical spacetimes but also of simpler quantum mechanical scenarios. Formulating a good notion of \enquote{event} is also a pertinent question in the literature on quantum foundations, as it plays an important role in several interpretations of quantum mechanics \cite{Wallace2012, Fuchs2013, Brukner2017, Adlam2023nonabs, ormrod2024quantum}, in the field of causal inference and, by extension to the quantum realm, in quantum causal models \cite{Costa_2016,barrett2020quantum,Barrett2021rev} and indefinite causal order \cite{Hardy_2005,Chiribella_2013, Oreshkov_2012, Oreshkov_2019, Paunkovic_2020, Faleiro2023operational, Vilasini_2022, Ormrod2023causal, delahamette2022quantum}. 
What constitutes an event varies significantly across these different contexts. Given the multitude of different notions of event, we will not fix a definition at this point but rather consider several examples illustrating what an event is \emph{not}. Specifically, we will identify which properties may be problematic for defining an event in the quantum context. In the course of this, we will encounter a varied spectrum of events, ranging from the application of an operation, to a coincidence of systems, to a change in a system's state. By collecting and combining these clues, we will eventually arrive at an \emph{operational} notion of event that depends on the measurement context and the phenomenon under consideration. More concretely, we will argue that a property of an event should be regarded as inherent if one can, in principle and without destroying the phenomenon of interest, access the property.\\

\textit{The Atomic Decay Clue.} As a first example, consider the decay of an atom of, for example, Radium-226. This event is characterised by the transition from Radium-226 to Radon-222 and an alpha-particle. If the system is isolated and no observation takes place, it will end up in a superposition of having and not having decayed. Consequently, there is a non-vanishing probability for the decay to occur at any given time $t_\ast$. Should we therefore say that there is an event \enquote{the atom decays at time $t_\ast$}? While one might first be inclined to do so, this clashes with the intuition that the atom really only decays once -- albeit, in a superposition of different times. This can be taken as evidence that a more sensible notion of event in this context is given by \enquote{the atom decays (at some point in time)}. We can make this intuition more precise. While \enquote{the atom decays at time $t_\ast$} can become an actual event if measured by detecting the decay at time $t_\ast$ (e.g.~by activating a Geiger counter at time $t_\ast$), in this case the superposed state is replaced by an updated state after the measurement. If, on the other hand, we wanted to verify that the state of the atom is indeed in a superposition of different decay times (e.g.~through an appropriate interference experiment), operationally accessing the decay time would disturb the verification procedure. This leads to a coarse-grained notion of event that captures the common features of the fine-grained events \enquote{the atom decays at time $t_\ast$} while excluding any reference to non-accessible attributes. Moreover, unlike \enquote{the atom decays at time $t_\ast$}, this notion does not lead to an infinite number of events and thus, if included in the theory's ontology, avoids an excessive number of distinct entities. Overall, we see that, depending on the measurement context, the time at which the event happens may or may not be regarded as a defining property thereof.\\

\textit{The Double-Slit Clue.} As a second example, let us return to the beginning of everyone's first lecture on quantum mechanics and consider the double-slit experiment \cite{FeynmanVol3}: A coherent beam of photons impinges on two closely spaced slits, producing an interference pattern on a screen placed behind the slits. The behaviour of a single photon is commonly described as passing through the double-slit in a superposition of the left and the right slit, respectively. Does this passing through the slits constitute one single or two separate events? Again, we are faced with the question of which properties to take into account when discriminating between the events. Clearly, most people would agree that the photon only passes through the double-slit once -- albeit in a superposition of spatial trajectories. Nevertheless, it also seems valid to consider the event \enquote{the photon passes through the left slit} as distinct from the event \enquote{the photon passes through the right slit}. However, from an operational perspective, such an assertion seems problematic, for it would require being able to \emph{measure}, in principle, which of the two slits the photon actually passes through. Any such measurement would inevitably disrupt the interference pattern. Thus, it is impossible to operationally distinguish the above two events without destroying the phenomenon we were studying in the first place. If instead one characterises the event simply as \enquote{the photon passes through the double-slit}, without specifying \emph{which} slit it went through, the relevant properties of this coarse-grained event \emph{can} be accessed operationally without altering the interference pattern. For example, one could perform a delayed-choice quantum-erasure experiment \cite{Kim2000} to confirm that the photon went through \emph{a} slit, without any effects on the pattern created on the screen. 

In keeping with the theme of this essay, let us now consider, as a gedankenexperiment, a gravitational analogue of the original double-slit setup. Take a massive object in superposition of two locations, left and right, and let a photon travel freely in the vicinity of this gravitational source. Taking the gravitational field to be in superposition, the photon's trajectory will branch into two paths, entangled with the location of the massive object \cite{delaHamette2021falling}. At its closest distance to the respective location of the massive object, the photon is reflected by a mirror and sent on two converging paths.
If we now recombine the mass adiabatically, we can interfere the photon with itself and produce an interference pattern, as in the optical analogue. Let us take a closer look at the event \enquote{the photon impinges on a mirror}. Since there are two different mirrors, it may seem natural to define two distinct events, \enquote{the photon impinges on the left mirror} and  \enquote{the photon impinges on the right mirror}, in superposition. However, just as in the optical case, we have no operational access to the property distinguishing the two events. If we could measure \emph{which} mirror the photon hits, at least in principle, we would acquire which-path information and thus destroy the interference. What we can measure is \emph{that} the photon hits one of the mirrors, i.e.~the event \enquote{the photon impinges on a mirror}. Thus, from an operational standpoint and within the context of this phenomenon, it seems that the latter notion of event is more appropriate. One can, of course, uphold the position that nevertheless the events are distinct. However, one then has to accept that they are distinguished by a property that cannot be operationally accessed within the context of the phenomenon that is being described. If the latter is changed, other measurements may become available, turning the spatial location into an operationally accessible quantity. Overall, we conclude that the spatial location at which the event happens may or may not be regarded as a defining property thereof, depending on the context.\\

\textit{The Quantum Switch Clue.} As a third clue, we want to consider an example from the intersection of quantum foundations with quantum gravity: indefinite causal order (ICO) \cite{Hardy_2005,Chiribella_2013, Oreshkov_2012, Oreshkov_2019, Paunkovic_2020, Faleiro2023operational, Vilasini_2022, Ormrod2023causal, delahamette2022quantum}. This provides a broad playground to discuss several potential properties of events and is, in fact, an instance for which the definition of event has important repercussions.

\begin{figure}
    \centering
    \subfigure[Optical quantum switch.]
    {
        \centering
        \hfill
        \includegraphics[scale=0.25]{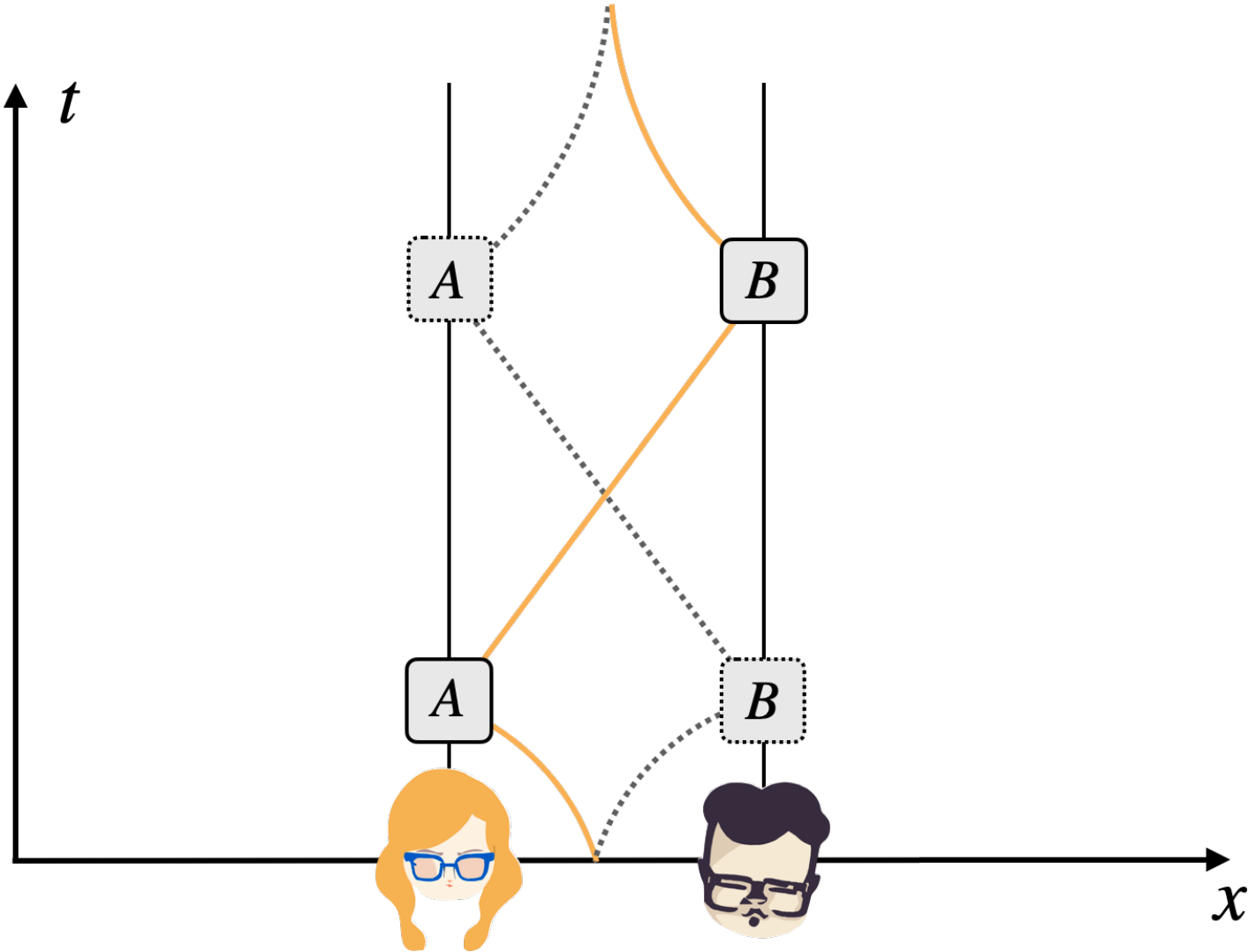}
        \label{fig:optical_switch}
    }
    \hspace{1.5cm}
    \subfigure[Gravitational quantum switch.]
    {
        \centering
        \hfill
        \includegraphics[scale=0.25]{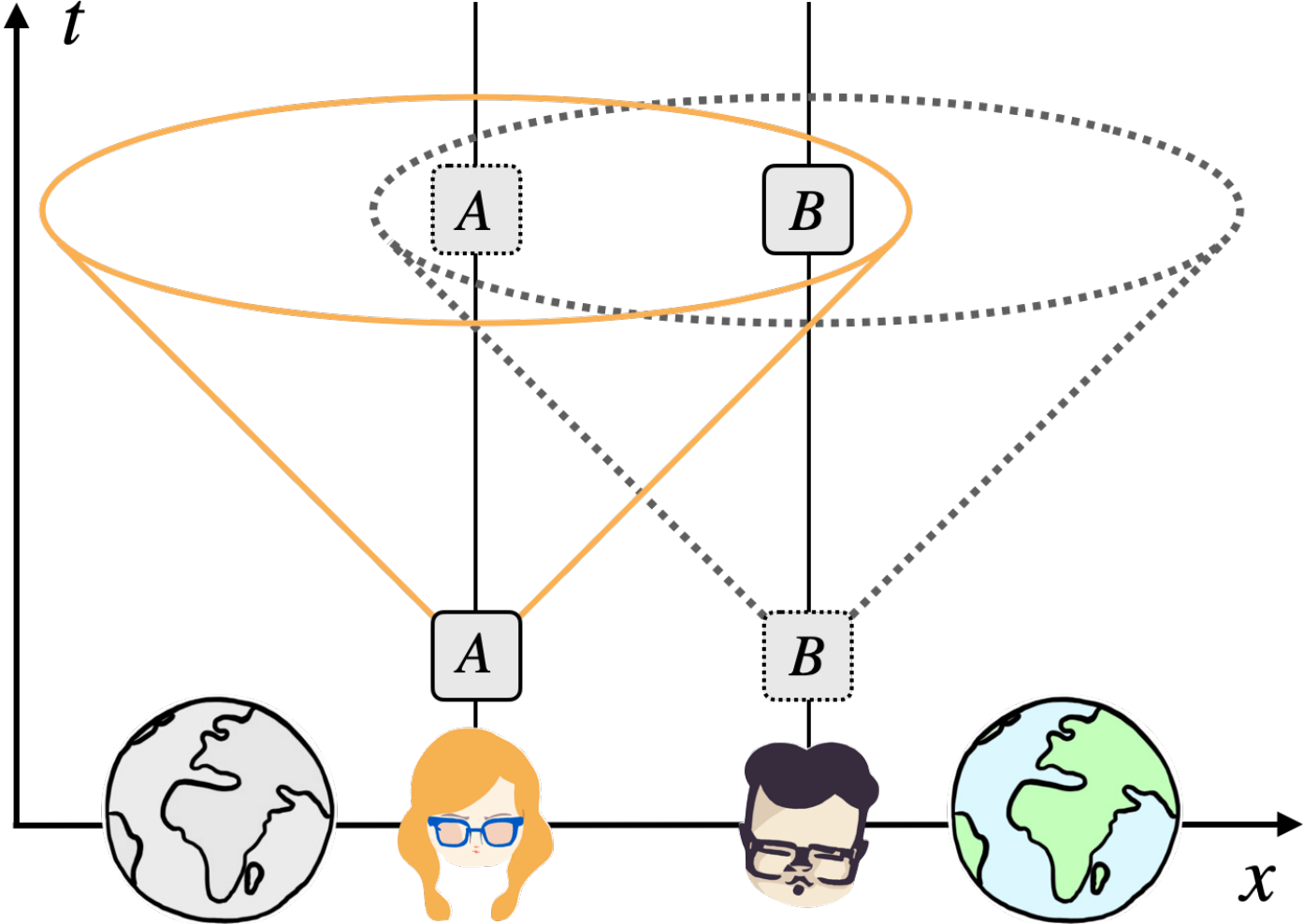}
         \label{fig:gravitational_switch}
    }
    \caption{\emph{Different implementations of the quantum switch.} Alice and Bob apply their operations $A$ and $B$ in a superposition of causal orders. While the indefinite causal order in (a) is created through a superposition of paths (represented by the orange, solid and the grey, dotted lines respectively), in (b) it is due to a superposition of the gravitational source (saturated vs.~grey Earth), which leads to a superposition of lightcones (orange, solid vs.~grey, dotted lines respectively).}
    \label{fig:quantum_switch}
\end{figure}

The study of ICO is too broad a field to fully do it justice in the context of this essay. In the following, we thus restrict to a few concrete implementations of ICO, most notably the quantum switch in its various incarnations \cite{Chiribella_2013, Oreshkov_2012, Procopio_2015, zych_2019}. In its simplest form, the quantum switch represents a process in which two operations, $A$ and $B$, are implemented in a quantum-controlled superposition of orders, i.e.~$A$ before $B$ and $B$ before $A$. In general, $A$ and $B$ are arbitrary completely positive maps (the most general quantum operations); we can, however, for the purpose of this essay, think of them as unitaries applied to a target system. Given an initial state $\ket{\psi_0}=\frac{1}{\sqrt{2}}(\ket{0}_C+\ket{1}_C)\ket{\phi}_T$ of the control and target system, the quantum switch realises the state $\ket{\psi_1}=\frac{1}{\sqrt{2}}(\ket{0}_C\otimes BA\ket{\phi}_T+\ket{1}_C\otimes AB\ket{\phi}_T)$. There are various different implementations of the quantum switch. The first is known as the \emph{optical quantum switch} and has been realised in the laboratory \cite{Procopio_2015, Rubino_2017, Goswami_2018}. It involves a photon as the target system travelling in a superposition of paths through two different waveplates, thereby implementing the unitaries $A$ and $B$ respectively in a quantum-controlled superposition of orders. At the end, the photon paths are re-interfered and ICO is verified using a causal witness measurement. Fig.~\ref{fig:optical_switch} represents this protocol in a spacetime diagram.
One can clearly see that operation $A$ is applied to the target system in a delocalised manner, that is, in a superposition of two different spacetime points. The same considerations apply to operation $B$. One may be tempted to say that there are four different events involved in the optical quantum switch. Importantly, these events are characterised by the application of a specific operation at a specific \emph{spacetime point}. While this view of the optical switch as a 4-event process is endorsed by a number of people \cite{Paunkovic_2020, Faleiro2023operational, Ormrod2023causal}, others believe that there are only two events involved in this process \cite{Araujo_2014, Procopio_2015, Oreshkov_2019, delahamette2022quantum}.\footnote{The two sides of this debate are referred to as the \enquote{spatio-temporalist} vs.~the \enquote{dynamicist} in \cite{Ormrod2023causal}. Vilasini and Renner also analyse this debate, distinguishing between more coarse-grained and fine-grained processes \cite{Vilasini_2022}.} In this coarse-grained view, events are characterised solely by the application of operation $A$ and $B$, respectively. While they happen in a (space)time-delocalised manner, each operation is still only being applied once -- just as the particle in the double-slit experiment only passes through the double-slit once. 
Moreover, just as one cannot access which slit the photon passes through, one cannot operationally access the spacetime locations of the events without destroying the interference, supporting the coarse-grained view.\footnote{While protocols to measure the number of spacetime points in various implementations of the quantum switch have been proposed \cite{Paunkovic_2020}, we believe that the measurement procedure that does not decohere the target system and destroy the interference pattern of the quantum switch cannot distinguish between two and four involved spacetime points with certainty in a single run of the experiment, as is claimed in \cite{Paunkovic_2020}. In particular, one can construct scenarios that involve four spacetime points for which their protocol would imply two, and vice-versa, in some runs of the experiment.}
In short, the spacetime location is not regarded as a relevant property of the event in this view. The question of how many events occur in the quantum switch has concrete consequences for the significance of the performed experiments. This is because causal relations are defined \emph{between events}. If there are four different events, the optical quantum switch merely implements a superposition of two definite causal orders between different pairs of events. If, on the other hand, there are only two events $E_A$ and $E_B$, we truly have a superposition of opposite causal orders, i.e.~$E_A$ before $E_B$ and $E_B$ before $E_A$. As a result, there is an ongoing debate as to whether the optical quantum switch is merely a simulation or a true realisation of ICO.

There is less disagreement when it comes to the \emph{gravitational} quantum switch \cite{zych_2019}. In this thought experiment, a large massive object is coherently superposed in two locations, sourcing a gravitational field in a superposition of two semiclassical configurations. A probe particle is sent along a superposition of trajectories, entangled with the position of the massive object, and passes through the laboratories of two agents, Alice and Bob, once in each of the branches. The setup is tuned such that, upon the passing of the target system through each laboratory, the local clock of the respective agent shows proper time $\tau_\ast$. Alice and Bob each apply their operations, $A$ and $B$, at precisely this time. Due to the superposition of gravitational fields, this happens in a superposition of opposite orders, thus implementing the quantum switch. Since in this setup it can be ensured that the target system only passes through each laboratory once, it is generally agreed that each operation is only applied once, resulting in a total of two events. Therefore, the gravitational switch is widely considered a proper realisation of ICO. When depicted in a spacetime diagram (see Fig.~\ref{fig:gravitational_switch}), however, one might come to believe that this scenario involves four events, after all. This is because, relative to the time measured by a far away observer, the probe passes through Alice's and Bob's laboratory in a superposition of spacetime coordinates. The assertion that there are four events not only contradicts the previously established intuition that the system really passes through each laboratory only once, but is problematic for a deeper reason. Since general relativity is a diffeomorphism-invariant theory, coordinate positions do not have intrinsic meaning of their own. By applying different coordinate transformations in each of the branches, one can always localise Alice's application of her operation to a single spacetime point, and similarly for Bob. In other words, one can always map a 4-point switch into a 2-point switch by choosing an appropriate quantum coordinate system \cite{delahamette2022quantum}. In addition to the previous discussion on being able to operationally access any inherent property without destroying the phenomenon of interest, we take this to imply that the spacetime location of an event should not be regarded as an inherent property, since it is a frame-dependent quantity \cite{kabel2024identification}.

Lastly, let us consider a variation of the quantum switch, which introduces a further dimension to our discourse on the notion of event. Returning to the abstract setup of the switch with two time-delocalised operations $A$ and $B$, let us now assume that, in the first branch, an operation $A_1$ is applied at time $t_1$ while a different operation $A_2$ is applied at time $t_2$ in the second (and similarly for $B_1$ and $B_2$).
This modification of the switch is referred to as the \emph{routed quantum switch} in \cite{Ormrod2023causal}. It presents us with a more intricate problem: can we still consider the event of applying operation $A_1$ as the same event as applying operation $A_2$? At first sight, answering this in the affirmative would seem like we have gone too far. For surely, the application of two different operations must instantiate two different events! Yet, if we take seriously the operational account of an event that has become clear throughout the previous examples, this intuition must be questioned. Just as we cannot access the time at which the atom decayed, or which slit the photon passed through, or the exact spacetime location of the events in the quantum switch, without destroying the phenomenon of interest, the same limitation applies to the routed quantum switch. Here, accessing which operation is applied to the photon would also destroy the interference. This is true even if it can be operationally verified that operation $A_1$ is applied at time $t_1$ and operation $A_2$ at time $t_2$. Since the photon arrives in a superposition of times in each lab, there is nevertheless no information about which of the two operations will be applied to the photon. This raises the critical question: How do we then define an event within the context of the routed switch? Having removed this much structure from the notion of event, what remains? The answer depends on the concrete implementation of the experiment and which attributes are still accessible. In an optical implementation, the unitary operations can be applied to the target system using waveplates, which are now taken to rotate in time such that different unitaries are applied at different times. In this case, an event would be characterised by the photon passing through one particular waveplate, say, that marked with label $A$ or $B$, but not by the specific rotation angle of the plate.
In a more interventionist picture, it would be the fact that a particular agent applies \emph{an} operation on the photon or that an operation is applied to the photon in a particular laboratory. 
Importantly, whatever property we use to define and distinguish events in the quantum context, we should ensure that we can, in principle, access this property without destroying the phenomenon of interest.\\

\textit{Putting the Pieces Together.} 
This also applies to the context of spacetimes in superposition, which we considered in the introduction. Recall the thought experiment in which two satellites collide in a superposition of spacetimes. Suppose we want to conduct an interference experiment to experimentally verify the superposition. By making use of our extensive experimental capabilities, we re-interfere both the gravitational source and the satellite remnants and measure the resulting interference pattern. Let us now revisit the questions raised at the start: is it still the \emph{same} event if the satellites collide at different spacetime points, velocities, angles, or forces? As for the spacetime location of the events, we have seen that it is a frame-dependent property and thus depends on the quantum coordinate system that one chooses \cite{kabel2024identification}. Moreover, just like measuring other potential properties such as velocity, angle, or force that differ across the branches of the superposition, measuring explicitly the concrete spacetime location of the event (relative to the experimental setup) would allow us to acquire which-path information and thus destroy the interference pattern. Even when there is an explosion in one branch and a non-explosive crash in the other, we have to take seriously that we cannot operationally access the property of whether the explosion occurred without disturbing the recombination process.\footnote{Of course, given our current technological capabilities, decoherence would be unavoidable; however, if one maintains complete control over all emitted photons to prevent any which-branch information from being transmitted to the environment, e.g.~by enclosing the setup within a sealed capsule, one could, in principle, recombine the branches.} Like the claim that the application of two different operations $A_1$ and $A_2$ might, in some contexts, constitute the \emph{same} event, this might seem like a radical statement. However, we believe that it is a direct consequence of our operational understanding of the notion of event.

More precisely, the ingredients that went into our construction of this notion are the following: (i) the \emph{Identity of Indiscernibles} \cite{Leibniz_1956, sep-identity-indiscernible}, that is, the metaphysical stance that if two entities are identical in all of their properties, one should not treat them as distinct, and (ii) what one might call \emph{$\text{Operationalism}^+$}. By operationalism, we mean the positivist stance that one should limit oneself to only those properties that are operationally accessible, that is, can be measured or observed. $\text{Operationalism}^+$ takes this further by restricting to quantities that can be measured or observed, in principle, \emph{within the given context}, that is, without disrupting the phenomenon of interest.\footnote{While the term $\text{Operationalism}^+$ is not used in any of these works, we believe that this stance is similar to the views purported in \cite{Bohr_1931, Bub_2017, Healey_2017, Janas_2022, Cuffaro2023}.} While this might seem like a drastic extension of operationalism, we believe it naturally takes into account that, in quantum theory, a measurement is a constitutive element of the phenomenon under investigation --  measuring one observable can affect the empirical outcomes of the measurement of another.

The resulting notion of event has pertinent implications for several areas in the foundations of physics. Firstly, events play an important role as the relata of causal relations \cite{Lewis_1986, Pearl_2009} and thereby in the fields of (quantum) causality and (quantum) causal modelling \cite{Costa_2016,barrett2020quantum, Barrett2021rev}. Thus, what we consider to be distinct events has significant implications for the causal structure. This became evident in the context of the quantum switch, where one obtains definite or indefinite causal order, depending on how one identifies the different causal relata across the superposition \cite{Ormrod2023causal}. We have seen that the spacetime location of an event cannot be accessed operationally, since any such measurement would decohere the quantum switch. Thus, we conclude that the location in spacetime is not to be regarded as an inherent property and the quantum switch represents a two-event process with ICO. More generally, as shown in \cite{Vilasini_2022}, causal processes have more or less fine-grained representations. The stance of the authors of \cite{Vilasini_2022} can be read as follows: if a fine-grained representation with acyclic causal structure (i.e.~a causal order that can be represented by a graph without loops) on a fixed spacetime background exists, then this process ought to be characterised as having definite causal order. In this essay, we have challenged this view by arguing that there are setups in which a coarse-grained notion of event is more appropriate than its fine-grained counterpart, on the grounds that it is impossible to operationally access the fine-grained structure.

Moreover, understanding in detail what an event is, and is not, may help to clarify the conceptual underpinnings of relational quantum mechanics (RQM), for it relies heavily, at least in some formulations \cite{Rovelli1996, LaudisaRovelli2021, AdlamRovelli2023}, on the notion of an interaction or event. In particular, the ontology suggested in recent work is a \enquote{flash-ontology} of sparse events or interactions \cite{Rovelli2018, LaudisaRovelli2021, AdlamRovelli2023}. However, the notion of an event as arising in an interaction whereby the systems involved take on definite values relative to one another \cite[p.~11]{AdlamRovelli2023} is quite vague. We thus believe that our analysis of which properties can or cannot constitute an event can help to fill this gap. While RQM, where events are a fundamental part of the ontology, is a particularly relevant arena for the notion of event, it also plays a considerable role in other interpretations. For instance, the question of what constitutes different events is intimately connected with the notion of branching in many-worlds interpretations \cite{sep-qm-manyworlds}. In particular, depending on \enquote{when} or under what conditions branching occurs, one would presumably end up with different numbers of distinct events. It would be an interesting question for future research to investigate whether there is a reading of the many-worlds interpretation that is consistent with our operationalist notion.
Finally, our understanding of event appears to align well with recent works on the non-absoluteness of events in quantum theory \cite{Fuchs2013, Brukner2017, Bong2020, Di_Biagio_2021, Moreno2022eventsinquantum, Brukner2022, ormrod2022nogo, Adlam2023nonabs, ormrod2024quantum}. There, events are not absolute but relative to observers, measurement contexts, (sets of) systems, or \enquote{bubbles}. While we believe that there is an overlap in spirit between our understanding of events and these works, they differ in what concretely events are relative to.

The observation that events (or measurement outcomes or facts) are relative to certain \enquote{bubbles} also plays an important role in attempts to resolve the Wigner's friend paradox and extensions thereof \cite{Wigner_1962, Brukner2017, Brukner_2018, Frauchiger_2018, Cavalcanti_2021}. With our notion of events, we can make this observation more tangible. Since the latter depends on the measurement context, Wigner and the Friend should give different accounts of the events involved in this scenario. Relative to the Friend's measurement setup, who measures a spin-$1/2$ particle in her laboratory, it is possible to access her outcome -- the spin of the particle -- without disturbing her setup. Relative to Wigner's setup, in which the coherence of the entangled state of the Friend and the system is verified through an interference experiment, however, the property of the spin-$1/2$ particle is not accessible. This is because measuring the spin would disturb the interference pattern. Thus, the spin of the particle should not be taken as an inherent property of the event, relative to Wigner's experimental context or \enquote{bubble}. For Wigner, it is not meaningful to differentiate between events \enquote{spin up} or \enquote{spin down} -- just as it is not meaningful to talk about \enquote{the photon passing through the right slit} and \enquote{the photon passing through the left slit} in the double-slit experiment. Similarly, it is meaningless for Wigner to distinguish between \enquote{Friend sees spin up} and \enquote{Friend sees spin down}. An event that \emph{can} be defined meaningfully relative to Wigner's setup is that the Friend has measured \emph{an} event. This has been shown to be possible without decohering the entangled state of Friend and the system \cite{Deutsch1985} and thus without destroying Wigner's experimental setup. 

We thus see that questioning the conceptual foundations of basic notions, such as what constitutes an event, helps to shed light on various problems in the foundations of physics and suggests ways to understand the underlying ontology. We believe that this is particularly true in the context of quantum gravity, where, as we have seen, the simple notion of an event as a spacetime point, will no longer be sufficient. In fact, even when considering events as physical coincidences, such as the crash of the satellites, there remains an ambiguity in which properties are inherent to the event. Thus, it is important to clarify how this notion can be generalised to non-classical spacetimes, such as semiclassical spacetimes in superposition. This is particularly true if one takes events to be part of the fundamental ontology of the theory. However, we believe that even an understanding of an effective notion of events that can be applied to these regimes delivers foundational insights that can guide us in the construction of a future theory of quantum gravity. 

\begin{acknowledgments}
We thank Jeremy Butterfield for inspiring discussions and for his comments on an earlier version of this essay. We also thank Nikola Paunkovic, V.~Vilasini, and Marko Vojinovic for helpful discussions. This research was funded in whole or in part by the Austrian Science Fund (FWF) [10.55776/F71] and [10.55776/COE1]. For open access purposes, the author has applied a CC BY public copyright license to any author accepted manuscript version arising from this submission. Funded by the European Union - NextGenerationEU. 
VK acknowledges support through a DOC Fellowship of the Austrian Academy of Sciences.
This publication was made possible through the financial support of the ID 62312 grant from the John Templeton Foundation, as part of The Quantum Information Structure of Spacetime (QISS) Project (qiss.fr). The opinions expressed in this publication are those of the authors and do not necessarily reflect the views of the John Templeton Foundation.\\

\end{acknowledgments}

\nocite{apsrev42Control}
\bibliography{bibliography}
\bibliographystyle{apsrev4-2.bst}

\end{document}